\documentclass[aps,prl,twocolumn,groupedaddress,showpacs]{revtex4}

\usepackage[dvips]{graphicx}
\usepackage{epsfig}
\usepackage{amssymb}
\usepackage{amsmath}
\usepackage{color}

\begin{document}


\title{Driving Rydberg-Rydberg transitions from a co-planar microwave waveguide}


\author{S. D. Hogan, J. A. Agner, and F. Merkt}

\affiliation{Laboratorium f\"ur Physikalische Chemie, ETH Z\"urich, CH-8093 Z\"urich, Switzerland}
\author{T. Thiele, S. Filipp, and A. Wallraff}
\affiliation{Department of Physics, ETH Z\"urich, CH-8093 Z\"urich, Switzerland}


\date{\today}

\begin{abstract}
The coherent interaction between ensembles of helium Rydberg atoms and microwave fields in the vicinity of a solid-state co-planar waveguide is reported. Rydberg-Rydberg transitions, at frequencies between 25~GHz and 38~GHz, have been studied for states with principal quantum numbers in the range 30 -- 35 by selective electric-field ionization. An experimental apparatus cooled to 100~K was used to reduce effects of blackbody radiation. Inhomogeneous, stray electric fields emanating from the surface of the waveguide have been characterized in frequency- and time-resolved measurements and coherence times of the Rydberg atoms on the order of 250~ns have been determined.
\end{abstract}

\pacs{}

\maketitle

Atoms and molecules in Rydberg states of high principal quantum number $n$ are very sensitive to low-frequency electromagnetic fields. This sensitivity arises because (i) the energy differences between states for which $|\Delta n|=1$ scale with $n^{-3}$, corresponding to frequencies in the millimeter-wave or microwave domains above $n=25$, and (ii) the dipole moments for these $|\Delta n|=1$ transitions scale with $n^{2}$ and are approximately equal to $n^{2}e\,a_0$. As a result, Rydberg atoms have been exploited extensively over the last three decades in studies of population transfer~\cite{beiting79,gallagher79} and ionization~\cite{spencer82,seiler11} by blackbody radiation. In the context of cavity quantum electrodynamics (QED), they have been used to study the modification of spontaneous emission rates of atoms surrounded by cavities~\cite{kleppner81,hulet85}, in the single-atom maser~\cite{meschede85}, in tests of the Jaynes-Cummings model~\cite{rempe87}, and in quantum non-demolition measurements of single photons confined in high-quality-factor microwave cavities~\cite{gleyzes07}.

Cavity QED effects have also been studied at microwave frequencies in solid-state superconducting circuits, e.g.,~using co-planar, superconducting microwave resonators coupled to superconducting qubits, in a variant of cavity QED known as circuit QED~\cite{wallraff04,schoelkopf08}. With the aim of eventually studying QED effects at microwave frequencies at a vacuum--solid-state boundary, we report here the coherent interaction of an ensemble of helium Rydberg atoms with microwave fields in a solid-state co-planar waveguide. Hybrid approaches of this kind to cavity QED are relevant to quantum information processing, where fast gate operations are expected from superconducting circuits and long-coherence-time quantum memories from Rydberg atoms~\cite{saffman10,comparat10}.

The possibility to carry out microwave spectroscopic measurements of Rydberg atoms or molecules located close to surfaces also opens up a new way to study Rydberg-atom--surface interactions. Of interest are, e.g.,~(i) interactions with the electric fields arising from surface adsorbates or patch potentials~\cite{neufeld11,carter11,tauschinsky10}, (ii) dipole-dipole interactions between Rydberg atoms and their image dipoles in the surface~\cite{anderson88,sandoghdar92}, and (iii) the prototypical process of charge transfer between a Rydberg atom and the conduction band of the surface leading to ionization~\cite{hill00,lloyd05,so11}.

The experimental method presented here is based on coherent coupling of Rydberg atoms to microwave fields in the vicinity of solid-state co-planar waveguides and allows the characterization of Rydberg-atom--surface interactions. This represents an initial step toward a new hybrid approach to cavity QED.
\begin{figure}
\includegraphics[width=7.0cm]{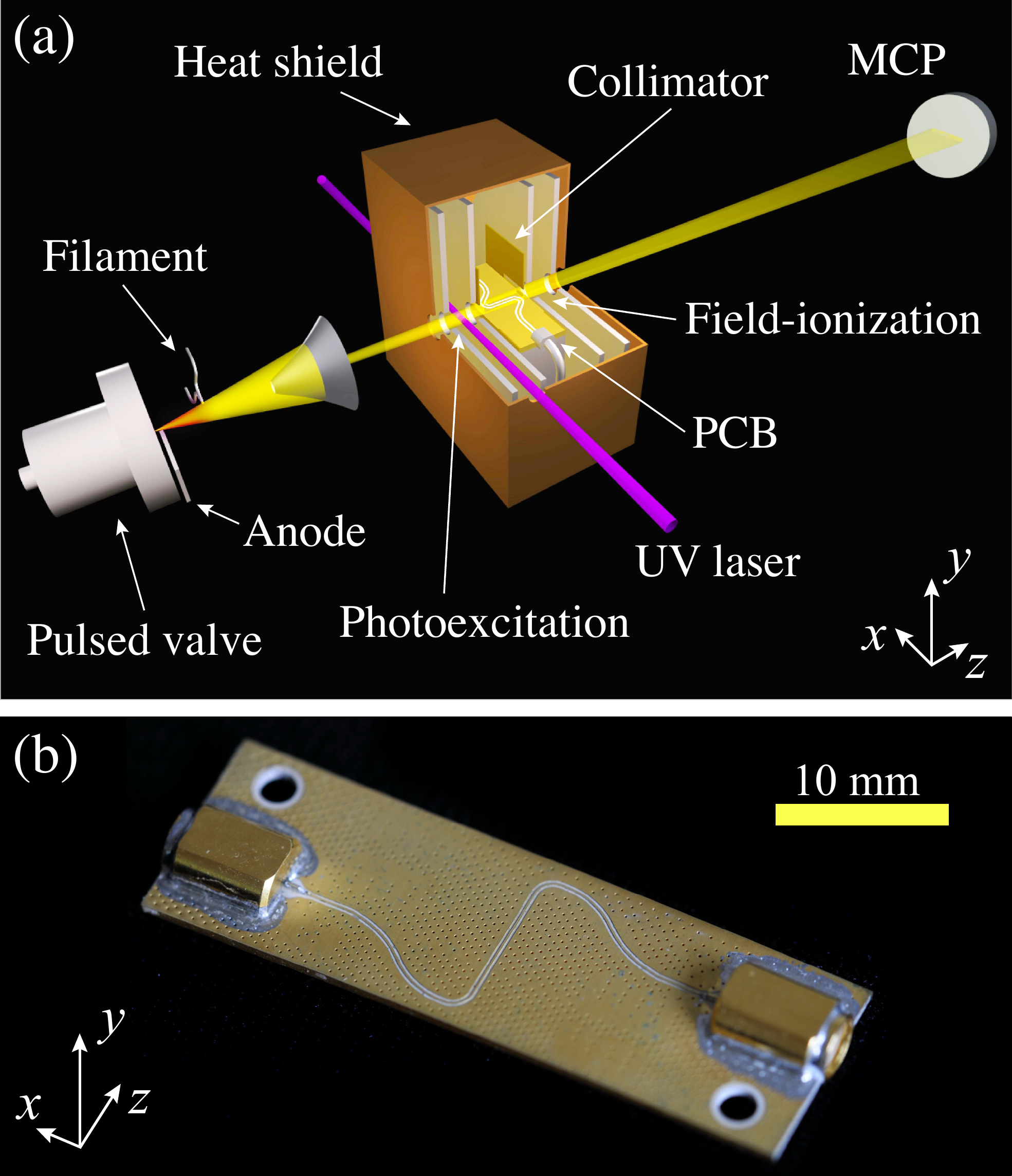}
\caption{(Color online) (a) Schematic diagram of the experimental apparatus (not to scale). (b) Photograph of the PCB containing the co-planar microwave waveguide.}
\label{fig1}
\end{figure}

A schematic diagram of the apparatus used in the experiments is presented in Fig.~\ref{fig1}(a). A pulsed supersonic beam of metastable helium atoms, propagating in the positive $z$-direction at a velocity of 1720~m\,s$^{-1}$, is prepared in an electrical discharge at the exit of a pulsed valve. The discharge takes place between a high-voltage anode (+250~V) positioned outside the valve and the grounded metallic poppet of the valve. A tungsten filament is used to seed the discharge with electrons, enhancing its stability and minimizing the translational temperature of the resulting beam~\cite{halfmann00}. The choice of helium atoms for these experiments is based on the desires to minimize changes in stray electric fields which can arise from the surface adsorption of other more polarizable atoms or molecules, and to avoid condensation upon cooling the central interaction region of the apparatus. 

After passing through a skimmer, the atomic beam enters the interaction region which is cooled to 100~K to reduce effects of blackbody radiation. This region contains two pairs of cylindrical electrodes aligned with the axis of the beam. A printed circuit board (PCB) [Fig.~\ref{fig1}(b)] with the co-planar microwave waveguide is located between these electrodes. Its surface is parallel to the beam propagation axis. At the midpoint between the first pair of electrodes in Fig.~\ref{fig1}(a), singlet 1s2s\,$^1$S$_0$ helium atoms are excited to $n$p Rydberg states, in the vicinity of $n=30$, in a single-photon transition at a wavelength of $\sim313$~nm using a frequency-doubled pulsed dye laser. 

\begin{figure}
\includegraphics[width=6.0cm]{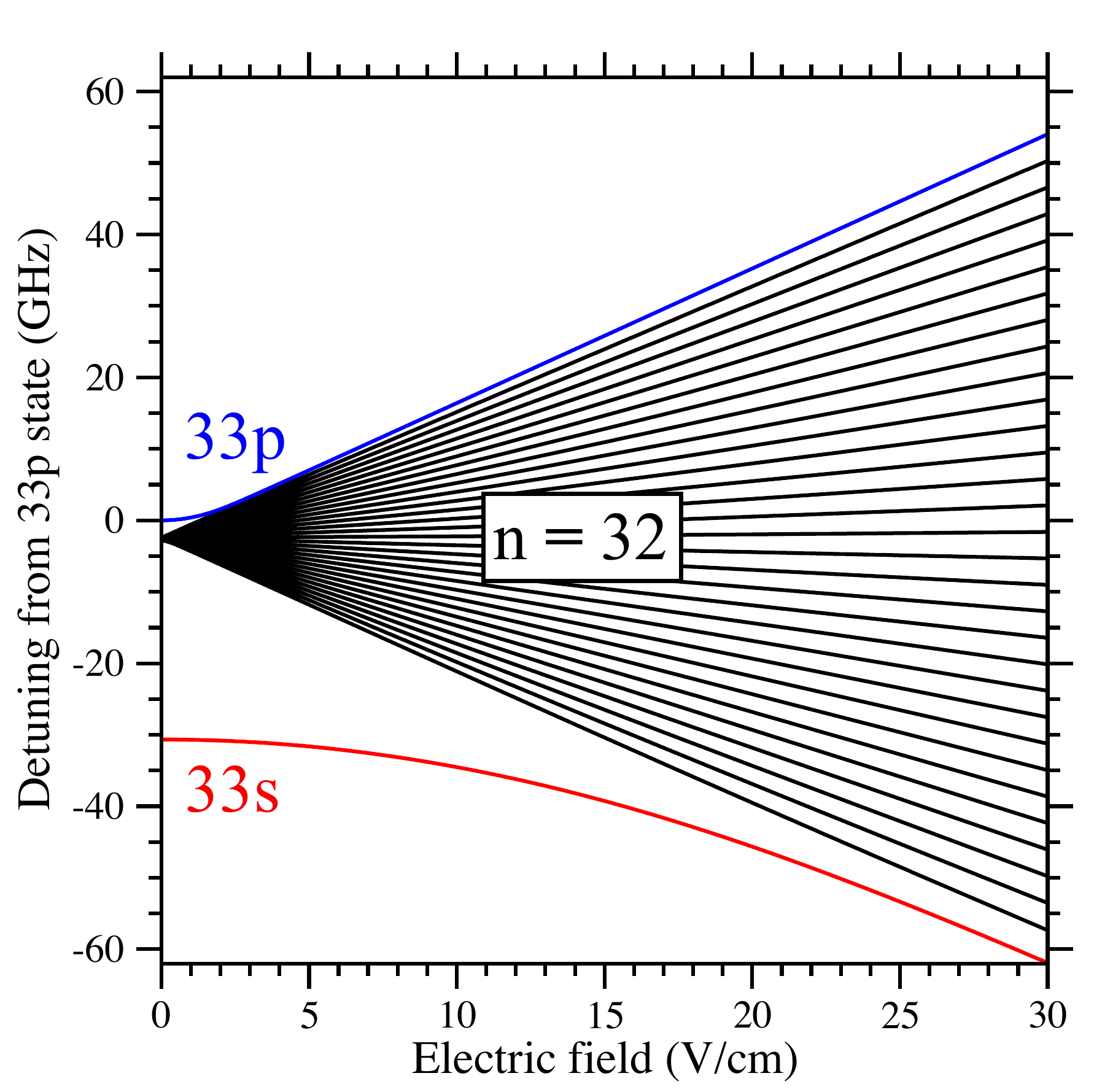}
\caption{(Color online) Stark map of singlet He Rydberg states in the vicinity of $n=32$.}
\label{fig2}
\end{figure}
The atomic beam is collimated by a pair of slits to ensure that the Rydberg atoms fly over the PCB within a distance of 1~mm from the surface in the $y$ dimension. When the atoms reach the center of the PCB, microwave pulses are applied to the co-planar waveguide to drive $n$p$\;\rightarrow\, n$s transitions. The co-planar waveguide has a 160~$\mu$m wide center conductor separated by 100~$\mu$m from the grounded outer region of the PCB. The 30~$\mu$m thick gold-coated copper surface layer of the PCB lies on top of a substrate with a relative permittivity of $\varepsilon_\mathrm{r}=10$. Upon reaching the second pair of electrodes, the Rydberg atoms are detected by state-selective electric-field ionization using a slowly rising ($\sim100$~V\,cm\,$^{-1}\mu$s$^{-1}$) electric-field pulse. The resulting electrons are extracted toward a microchannel-plate (MCP) detector.

The Stark effect in the vicinity of the singlet $n=32$ manifold of He is depicted in Fig.~\ref{fig2} in the form of a Stark map. This map was calculated by matrix diagonalization in a spherical $|n\,\ell\rangle$ basis, where $\ell$ is the orbital angular momentum quantum number of the Rydberg electron. Quantum defects of 1.1397, 0.9879, 0.0021 and 0.0004 were used for the singlet s, p, d and f Rydberg states of He~\cite{martin87} and the relevant radial integrals were calculated using the Numerov method~\cite{zimmerman79}. In the absence of external fields, the dipole-allowed microwave transition between the 33p and the energetically lower-lying 33s singlet Rydberg states of He occurs at a frequency of 30.66~GHz. The Stark shifts of transitions of this kind can be used as sensitive probes of stray electric fields~\cite{osterwalder99}.

\begin{figure}
\includegraphics[width=8.0cm]{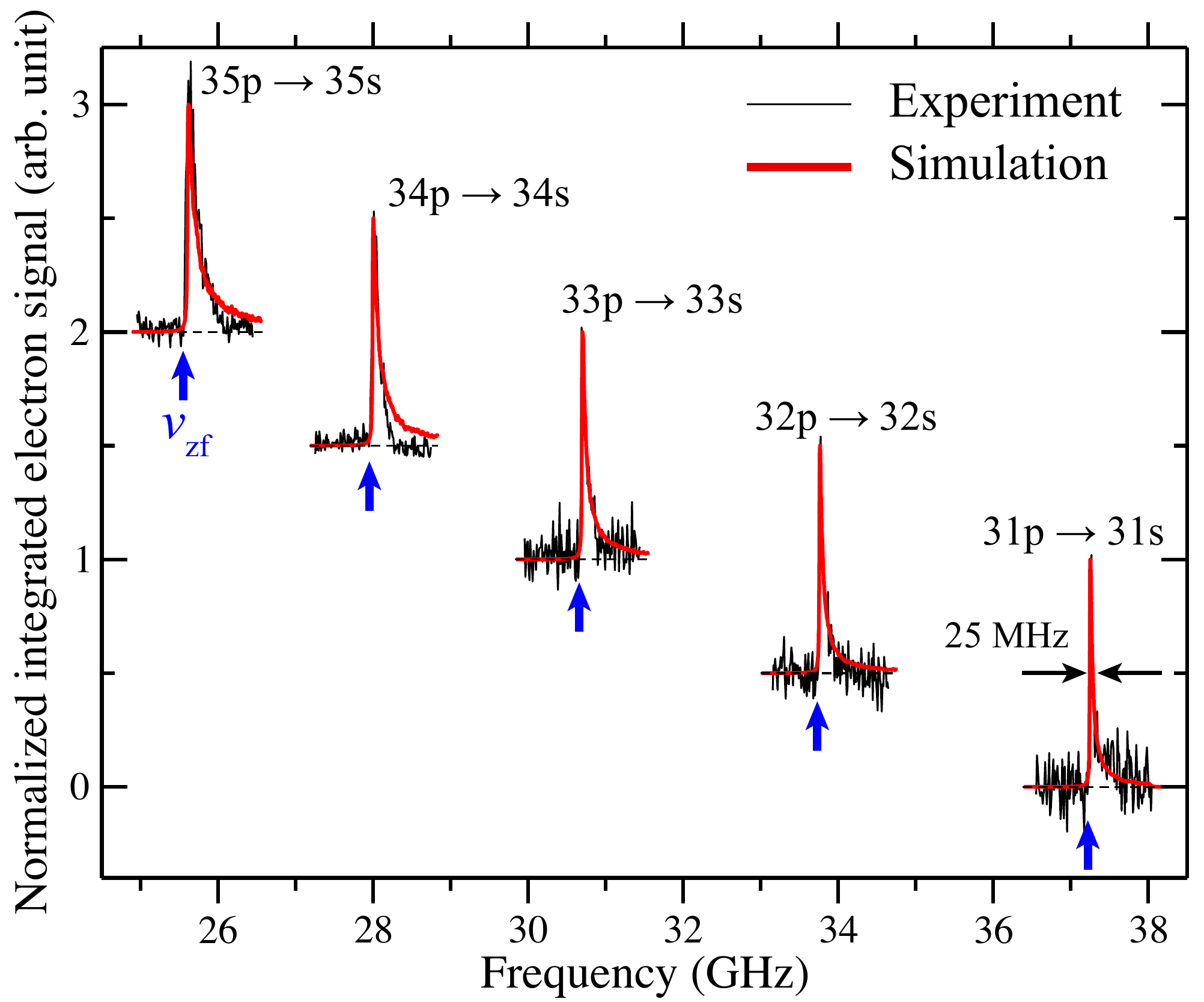}
\caption{(Color online) Microwave spectra of singlet $n$p$\;\rightarrow\, n$s transitions, in the range $n=31-35$, for He atoms traversing the co-planar waveguide. Simulated lineshapes (red) are overlaid on the experimental data with the zero-field transition frequencies, $\nu_{\mathrm {zf}}$, indicated by arrows.}
\label{fig3}
\end{figure}
To determine the magnitude of stray electric fields in the region close to the co-planar waveguide a set of pulsed microwave measurements, presented in Fig.~\ref{fig3}, were performed for different values of $n$. These measurements involved driving $n$p$\;\rightarrow\, n$s transitions for states between $n=31$ and $n=35$ and selectively detecting the final state. Microwave pulses of 1~$\mu$s duration were applied to the waveguide 6~$\mu$s after Rydberg photoexcitation, when the Rydberg atoms were located at the midpoint of the PCB in the $z$-dimension. A constant output power of the microwave source of $P_{\mu}=320$~nW was maintained throughout these measurements, with an attenuation of $\sim15$~dB between the source and the PCB. The spectra recorded exhibit strongly asymmetric line shapes with a sharp rise on the low-frequency side near the zero-field transition frequency,~$\nu_{\mathrm{zf}}$, and a long tail on the high-frequency side. These features, together with the strong $n$ dependence of the transition linewidths, (the narrowest line at the lowest $n$ has a full width at half maximum (FWHM) of 25~MHz), betray the presence of inhomogeneous stray electric fields~\cite{osterwalder99}. These fields are likely to arise from surface imperfections including patches and adsorbates. Such stray fields have been studied previously, and recent work has indicated that for distances larger than the patch size, the field strength scales with the inverse square of the atom-surface distance~\cite{carter11}. 

To elucidate the effect of these stray electric fields on the line shapes in the spectra in Fig.~\ref{fig3}, a numerical simulation was performed in which the primary fit parameter was the magnitude of the stray electric field. This simulation involved the generation of an ensemble of atoms, with a spatial distribution matching that of the experiment. These atoms were randomly positioned in the $xy$ plane at the mid-point of the PCB in the $z$ dimension, with a normal distribution of positions in the $x$ dimension with a FWHM of 2~mm centered above the waveguide. In the $y$ dimension, the atoms were described by a normal distribution of the same FWHM which was cut off at a distance of 1~mm above the PCB to account for the collimation of the atomic beam. 

The stray electric field was estimated by fitting $F=\nolinebreak[4]F_0/y^2$ simultaneously to all spectra, which led to $F_0=\nolinebreak[4]2.8\pm0.1\times10^{-5}$~Vm. The $n$p$\;\rightarrow\, n$s transition frequencies were calculated using the Stark shifts determined by the matrix-diagonalization procedure. To account for the inhomogeneity of the microwave field distribution surrounding the waveguide, a calculation of this field was carried out in the $xy$ plane above the PCB using a finite-element method. The microwave field strength $E_{\mu}$ experienced by each atom, and the field-dependent $n$p$\;\rightarrow\, n$s transition dipole moment $D_{n\mathrm{p}\,n\mathrm{s}}(F)$, were used to determine the degree of saturation $S_{n\mathrm{p}\,n\mathrm{s}}=[D_{n\mathrm{p}\,n\mathrm{s}}(F)\,E_{\mu}/(\hbar\,\gamma)]^2$ of the transition, where $\gamma$ is the natural linewidth. The transition dipole moments were calculated by summing the contribution of each dipole-allowed $\Delta\ell=\pm1$ transition between the $\ell$-mixed $n$p and $n$s states in the presence of the stray field. A Lorentzian homogeneous line shape with a FWHM of $\gamma\,\sqrt{1+S_{n\mathrm{p}\,n\mathrm{s}}}$ (where $\gamma<10$~MHz) was associated with each atom. The simulated line profiles of an ensemble of 20\,000 atoms were generated by summing these Lorentzian functions (thick red lines in Fig.~\ref{fig3}) and are in good agreement with the experimental data. The sharp rise close to the zero-field transition frequencies corresponds to atoms at larger distances from the surface of the PCB in the $y$ dimension where the stray field and its gradient are smallest. The tail at higher frequencies arises from atoms closer to the surface with larger Stark shifts. 

\begin{figure}
\includegraphics[width=8.5cm]{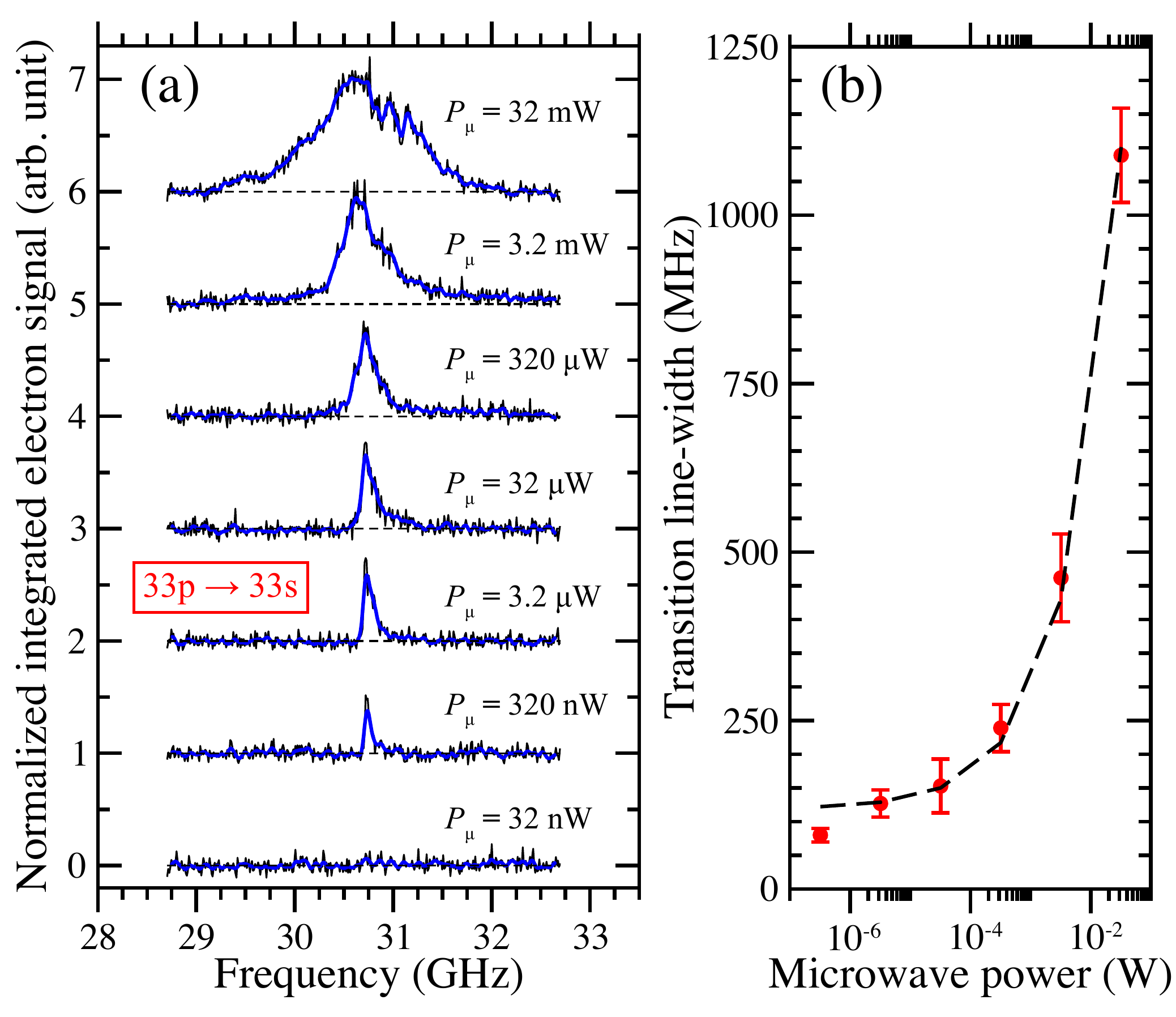}
\caption{(Color online) (a) Spectra of the singlet $33$p$\;\rightarrow\, 33$s transition in He as a function of microwave power. The output power of the microwave source, $P_{\mu}$, is indicated on the right-hand side of each spectrum. (b) Dependence of the measured line-width on $P_{\mu}$.}
\label{fig4}
\end{figure}
To investigate the degree of saturation that can be achieved for Rydberg-Rydberg transitions, a set of spectra of the $33$p$\;\rightarrow\, 33$s transition, presented in Fig.~\ref{fig4}(a), was recorded as a function of microwave power under otherwise identical conditions. Increasing the output power of the microwave source by a factors of 10 beginning with $P_{\mu}=32$~nW revealed that significant saturation could be achieved. The measured line widths range from 80~MHz at $P_{\mu}=320$~nW to 1.1~GHz at $P_{\mu}=32$~mW and are plotted as a function of microwave power in Fig.~\ref{fig4}(b). The experimental data has a $\sqrt{P_{\mu}}$ dependence as indicated by the fitted dashed curve representing the function $f(P_{\mu})/\mathrm{MHz}=119+5490\,\sqrt{P_{\mu}/\mathrm{W}}$. At low powers, inhomogeneous broadening by the stray electric field dominates, leading to a deviation from this behavior.

\begin{figure}
\includegraphics[width=8.5cm]{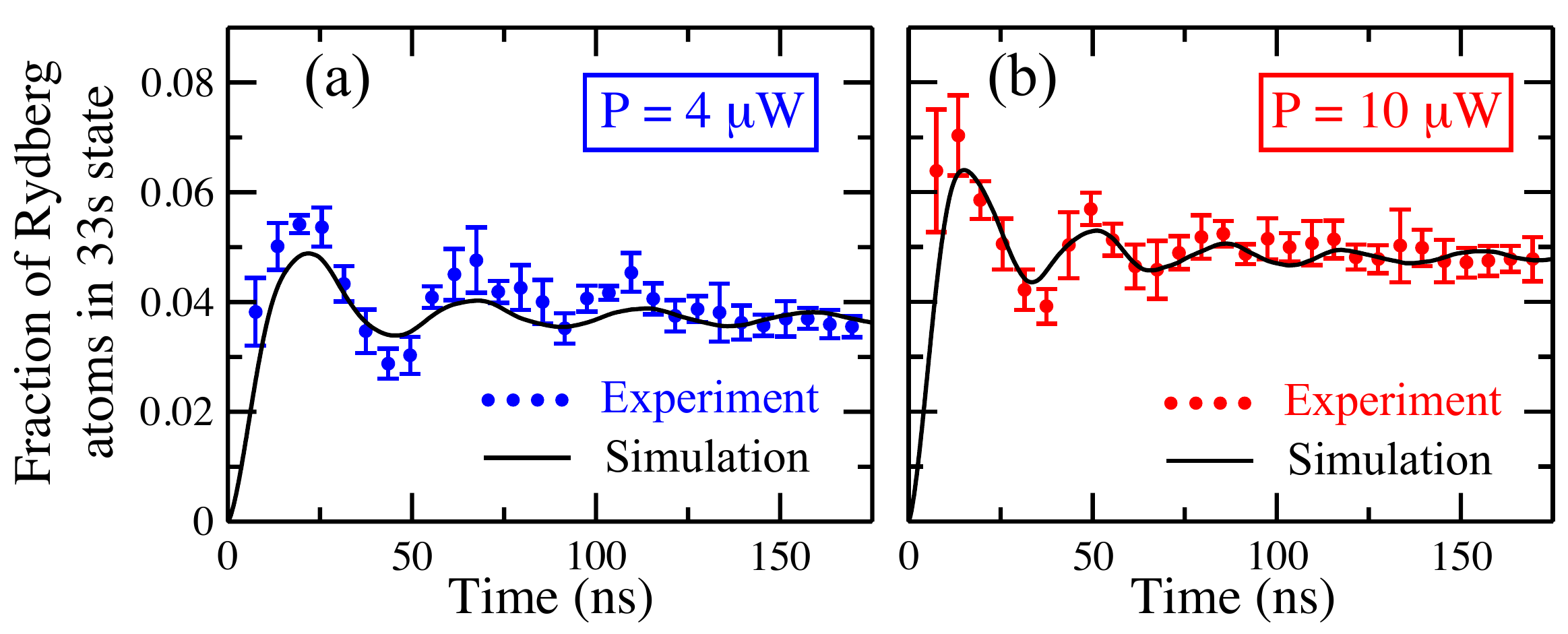}
\caption{(Color online) Fraction of Rydberg atoms in the 33s state as a function of microwave pulse duration for output powers of the microwave source of (a) 4~$\mu$W and (b) 10~$\mu$W. The solid black lines are the results of numerical simulations.}\label{fig5}
\end{figure}

The coherence of this atom-radiation interaction was investigated by selective detection of atoms in the 33s state as a function of microwave pulse duration at a fixed frequency and microwave power. The results obtained at a frequency of 30.76~GHz and microwave powers of 4~$\mu$W and 10~$\mu$W are presented in Fig.~\ref{fig5}. In each case, the experimental data are displayed with error bars indicating the standard deviation from the mean of six measurements. At early times, population is transferred from the initially excited 33p state to the 33s state. Rabi oscillations are then observed in the fraction of atoms in the 33s state. At later times, the contrast of the oscillations reduces because of ensemble dephasing and decoherence. Increasing the microwave power from 4~$\mu$W to 10~$\mu$W gives a corresponding increase in the observed oscillation frequency.

This coherent population transfer between the 33p and 33s states was simulated in an extension of the procedure described above. Since each atom experiences a different microwave field ($E_{\mu}$) and a different stray electric field ($F$ from Fig.~\ref{fig3}), individual Rabi oscillation frequencies, $\Omega_{33\mathrm{p}\,33\mathrm{s}} = D_{33\mathrm{p}\,33\mathrm{s}}(F)\,E_{\mu}/\hbar$, were determined. The fraction of the total ensemble of Rydberg atoms in the 33s state was then calculated by summing the contribution of each atom in the simulation. Because of the stray electric field each atom also experiences a different detuning $\Delta$. Consequently, the single-atom oscillation frequency becomes $\Omega'_{33\mathrm{p}\,33\mathrm{s}}=\sqrt{\Delta^{2}+\Omega^{2}_{33\mathrm{p}\,33\mathrm{s}}}$ and the contrast of the corresponding Rabi oscillations is reduced by a factor $(\Omega_{33\mathrm{p}\,33\mathrm{s}}/\Omega'_{33\mathrm{p}\,33\mathrm{s}})^2$. Across the ensemble, the oscillations therefore dephase with time.

Under the experimental conditions, $T_1$ is dominated by the fluorescence lifetime of the 33p state and is $>1$~$\mu$s. The motion of the atoms ($\sim1.7$~mm\,$\mu$s$^{-1}$) above the PCB  between regions of different microwave field and stray electric field gives rise to a time-dependence of $\Omega_{33\mathrm{p}\,33\mathrm{s}}$ and $\Delta$. This results in a decoherence time of $\sim$250~ns, which was determined from the experimental data, and led to the simulated black curves in Fig.~\ref{fig5}. In changing the microwave power from 4~$\mu$W in Fig.~\ref{fig5}(a) to 10~$\mu$W in Fig.~\ref{fig5}(b) the microwave field strength increases by a factor of 1.6. Best agreement was found between the simulations and the experimental data with a factor of $1.4\pm0.1$. The discrepancy is attributed to the limited precision with which the spatial distribution of the stray electric fields emanating from the surface of the PCB are modeled.

The coherent coupling of Rydberg atoms to microwave circuits demonstrated here opens up the possibility of exploiting hybrid atomic--solid-state systems of this kind in cavity QED and quantum information processing, including surface-based, non-destructive Rydberg atom detection. It is planned, in the next generations of experiments, to replace the normal metallic co-planar wave\-guide used in this work with a superconducting one and then to use a co-planar microwave resonator on a single-crystal substrate. The method of measuring stray electric fields presented here will be employed to select optimal materials for these devices. Further collimation combined with electrostatic guiding, and deceleration~\cite{hogan08a,seiler11} of the Rydberg atom beam will permit more precise control over the atom-surface distance and reduce decoherence caused by motion. In this way, the limit of single-atom--single-photon coupling will be approached.

\begin{acknowledgments}
We thank H. Schmutz, S. Marx, Ch. Gross and P. Allmendinger for help and discussions. This work is supported by the NCCR QSIT of the Swiss National Science Foundation, and the ERC under Projects 228286 and 240328. S.~F. thanks the Austrian Science Foundation for support.
\end{acknowledgments}

\end{document}